\documentclass[aps,prb,reprint,superscriptaddress]{revtex4-1}
\usepackage{graphicx,dcolumn,bm,xcolor,microtype,multirow,amscd,amsmath,amssymb,amsfonts,physics,wrapfig,txfonts,siunitx}
\usepackage[version=4]{mhchem}
\usepackage{comment}

\usepackage[
colorlinks=true,
linkcolor=blue,
citecolor=blue,
urlcolor=blue]{hyperref}

\newcommand{\br}{\boldsymbol{r}}
\newcommand{\dbr}{d\br}
\newcommand{\bx}{\boldsymbol{x}}
\newcommand{\dbx}{d\bx}
\newcommand{\Nel}{N}

\newcommand{\Occ}{O}
\newcommand{\Vir}{V}
\newcommand{\IS}{\lambda}

\newcommand{\T}[1]{#1^{\intercal}}

\newcommand{\singlet}{\uparrow\downarrow}
\newcommand{\triplet}{\uparrow\uparrow}

\newcommand{\Ec}{E_\text{c}}
\newcommand{\EHF}{E^\text{HF}}
\newcommand{\EcBSE}{E_\text{c}^\text{BSE}}
\newcommand{\EcTr}[1]{E_{\text{c},#1}^\text{Tr@BSE}}
\newcommand{\EcTrBSE}{E_\text{c}^\text{Tr@BSE}}
\newcommand{\EcACBSE}{E_\text{c}^\text{AC@BSE}}
\newcommand{\eps}{\epsilon}
\newcommand{\veps}{\varepsilon}

\newcommand{\Eg}{E_\text{g}}

\newcommand{\A}[2]{A_{#1}^{#2}}
\newcommand{\tA}[2]{\Tilde{A}_{#1}^{#2}}
\newcommand{\B}[2]{B_{#1}^{#2}}

\newcommand{\MO}[1]{\phi_{#1}}
\newcommand{\ERI}[2]{(#1|#2)}
\newcommand{\Om}[2]{\Omega_{#1}^{#2}}

\newcommand{\bO}{\boldsymbol{0}}
\newcommand{\bI}{\boldsymbol{1}}

\newcommand{\bA}[1]{\boldsymbol{A}^{#1}}
\newcommand{\btA}[1]{\Tilde{\boldsymbol{A}}^{#1}}
\newcommand{\bB}[1]{\boldsymbol{B}^{#1}}
\newcommand{\bX}[1]{\boldsymbol{X}^{#1}}
\newcommand{\bY}[1]{\boldsymbol{Y}^{#1}}
\newcommand{\bZ}[2]{\boldsymbol{V}_{#1}^{#2}}
\newcommand{\bK}{\boldsymbol{K}}
\newcommand{\bP}[1]{\boldsymbol{P}^{#1}}

\newcommand{\ie}{\textit{i.e.}}

\newcommand{\I}{\text{i}}
\newcommand{\bn}{\text{bn}}
\newcommand{\an}{\text{an}}

\begin{document}


\title{Scrutinizing $GW$-based methods using the Hubbard dimer}

\newcommand{\lcpq}{Laboratoire de Chimie et Physique Quantiques, Universit\'e de Toulouse, CNRS, UPS, France}
\newcommand{\lpt}{Laboratoire de Physique Th\'eorique, Universit\'e de Toulouse, CNRS, UPS, France}
\newcommand{\etsf}{European Theoretical Spectroscopy Facility (ETSF)}
\affiliation{\lcpq}
\affiliation{\lpt}
\affiliation{\etsf}
\author{Stefano Di Sabatino}
\affiliation{\lcpq}
\affiliation{\lpt}
\affiliation{\etsf}
%
\author{Pierre-Fran\c{c}ois Loos}
\affiliation{\lcpq}
\author{Pina Romaniello}
\affiliation{\lpt}
\affiliation{\etsf}

\begin{abstract}
Using the simple (symmetric) Hubbard dimer, we analyze some important features of the $GW$ approximation. We show that the problem of the existence of multiple quasiparticle solutions in the (perturbative) one-shot $GW$ method and its partially self-consistent version is solved by full self-consistency. We also analyze the neutral excitation spectrum using the Bethe-Salpeter equation (BSE) formalism within the standard $GW$ approximation and find, in particular, that i) some neutral excitation energies become complex when the electron-electron interaction $U$ increases, which can be traced back to the approximate nature of the $GW$ quasiparticle energies; ii) the BSE formalism yields accurate correlation energies over a wide range of $U$ when the trace (or plasmon) formula is employed; iii) the trace formula is sensitive to the occurrence of complex excitation energies (especially singlet), while the expression obtained from the adiabatic-connection fluctuation-dissipation theorem (ACFDT) is more stable (yet less accurate); iv) the trace formula has the correct behavior for weak (\ie, small $U$) interaction, unlike the ACFDT expression. 
\end{abstract}

\maketitle

\section{Introduction}
\label{sec:intro}
Many-body perturbation theory (MBPT) based on Green's functions is among the standard tools in condensed matter physics for the study of ground- and excited-state properties. \citep{Aryasetiawan_1998,Onida_2002,Martin_2016,Golze_2019} In particular, the $GW$ approximation \citep{Hedin_1965,Golze_2019} has become the method of choice for band-structure and photoemission calculations and, combined with the Bethe-Salpeter equation (BSE@$GW$) formalism, \citep{Salpeter_1951,Strinati_1988,Albrecht_1998,Rohlfing_1998,Benedict_1998,vanderHorst_1999,Blase_2018,Blase_2020} for optical spectra calculations. 
Thanks to efficient implementations, \citep{Duchemin_2019,Duchemin_2020,Duchemin_2021,Bruneval_2016,vanSetten_2013,Kaplan_2015, Kaplan_2016,Krause_2017,Caruso_2012,Caruso_2013,Caruso_2013a,Caruso_2013b,Wilhelm_2018} this toolkit is acquiring increasing popularity in the traditional quantum chemistry community, \citep{Rohlfing_1999a,Horst_1999,Puschnig_2002,Tiago_2003,Boulanger_2014,Jacquemin_2015a,Bruneval_2015,Jacquemin_2015b,Hirose_2015,Jacquemin_2017a,Jacquemin_2017b,Rangel_2017,Krause_2017,Gui_2018,Blase_2018,Liu_2020,Blase_2020,Holzer_2018a,Holzer_2018b,Loos_2020e} partially due to the similarity of the equation structure to that of the standard Hartree-Fock (HF) \citep{SzaboBook} or Kohn-Sham (KS) \citep{Hohenberg_1964,Kohn_1965} mean-field methods.  Several studies of the performance of various flavors of $GW$ in atomic and molecular systems are now present in the literature, \citep{Holm_PRB1998,Stan_2006,Stan_JCP2009,Blase_2011,Faber_2011,Bruneval_2012,Bruneval_2013,Bruneval_2015,Karlsson_PRB2016,Bruneval_2016, Bruneval_2016a,Boulanger_2014,Blase_2016,Li_2017,Hung_2016,Hung_2017,vanSetten_2015,vanSetten_2018,Ou_2016,Ou_2018,Faber_2014} providing a clearer picture of the \textit{pros} and \textit{cons} of this approach. There are, however, still some open issues, such as i) how to overcome the problem of multiple quasiparticle solutions, \citep{vanSetten_2015,Maggio_2017,Loos_2018,Veril_2018,Duchemin_2020,Loos_2020e} ii) what is the best way to calculate ground-state total energies, \citep{Casida_2005,Huix-Rotllant_2011,Caruso_2013,Casida_2016,Colonna_2014,Olsen_2014,Hellgren_2015,Holzer_2018b,Li_2019,Li_2020,Loos_2020e} and iii) what are the limits of the BSE in the simplification commonly used in the so-called Casida equations. \citep{Strinati_1988,Rohlfing_2000,Sottile_2003,Myohanen_2008,Ma_2009a,Ma_2009b,Romaniello_2009b,Sangalli_2011,Huix-Rotllant_2011,Sakkinen_2012,Zhang_2013,Rebolini_2016,Olevano_2019,Lettmann_2019,Loos_2020h,Authier_2020,Monino_2021}
In the present work, we address precisely these questions by using a very simple and exactly solvable model, the symmetric Hubbard dimer. Small Hubbard clusters are widely used test systems for the GW approximation \citep[e.g.][]{Verdozzi_1995,Schindlmayr_1998b,Pollehn_1998,vonFriesen_2010,Romaniello_2009a,Romaniello_2012}.  Despite its simplicity, the Hubbard dimer is able to capture lots of the underlying physics observed in more realistic systems, \citep{Romaniello_2009a,Romaniello_2012,Carrascal_2015,Carrascal_2018} such as, for example, the nature of the band-gap opening in strongly correlated systems as bulk NiO. \citep{DiSabatino_2016}
Here, we will use it to better understand some features of the $GW$ approximation and the BSE@$GW$ approach.  
Of course, care must be taken when extrapolating conclusions to realistic systems. 

The paper is organized as follows. Section \ref{sec:Theory} provides the key equations employed in MBPT to calculate removal and addition energies (or charged excitations), neutral (or optical) excitation energies, and ground-state correlation energies. 
In Sec.~\ref{sec:Results}, we present and discuss the results that we have obtained for the Hubbard dimer. We finally draw conclusions and perspectives in Sec.~\ref{sec:Conclusions}

\section{Theoretical framework}
\label{sec:Theory}
In the following we provide the key equations of MBPT \citep{Martin_2016} and, in particular, we discuss how one can calculate ground- and excited-state properties, namely removal and addition energies, spectral function, total energies, and neutral excitation energies. We use atomic units $\hbar=m=e=1$ and work at zero temperature throughout the paper.

\subsection{The $GW$ approximation}
\label{sec:GWA}
Within MBPT a prominent role is played by the one-body Green's function $G$ which has the following spectral representation in the frequency domain:
\begin{equation}
\label{Eqn:spectralG}
	G(\bx_1,\bx_2; \omega ) = \sum_\nu \frac{ \psi_\nu(\bx_1) \psi^*_\nu(\bx_2) }{ \omega - \eps_\nu + \I \eta \, \text{sgn}(\eps_\nu - \mu ) },
\end{equation}
where $\mu$ is the chemical potential, $\eta$ is a positive infinitesimal, $\eps_\nu = E_\nu^{\Nel+1} - E_0^{\Nel}$ for $\eps_\nu > \mu$, and $\eps_\nu = E_0^{\Nel} - E_i^{\Nel-1}$ for $\eps_\nu < \mu$.
Here, $E_\nu^{\Nel}$ is the total energy of the $\nu$th excited state of the $\Nel$-electron system ($\nu = 0$ being the ground state).
In the case of single-determinant many-body wave functions (such as HF or KS), the so-called Lehmann amplitudes $\psi_\nu(\bx)$ reduce to one-body orbitals and the poles of the Green's function $\eps_\nu$ to one-body orbital energies.

The one-body Green's function is a powerful quantity that contains a wealth of information about the physical system. In particular, as readily seen from Eq.~\eqref{Eqn:spectralG}, it has poles at the charged excitation energies of the system, which are proper addition/removal energies of the $\Nel$-electron system. Thus, one can also access the (photoemission) fundamental gap
\begin{equation}
\label{Eqn:EgFun}
	\Eg = I^{\Nel} - A^{\Nel},
\end{equation}
where $I^{\Nel} = E_0^{\Nel-1} - E_0^{\Nel}$ is the ionization potential and $A^{\Nel} = E_0^{\Nel} - E_0^{\Nel+1}$ is the electron affinity.
Moreover, one can straightforwardly obtain the spectral function, which is closely related to photoemission spectra, as
\begin{equation}
\label{Eqn:spectral_functions}
    A(\bx_1,\bx_2; \omega)=\frac{1}{\pi}\text{sgn}(\mu-\omega)\Im G(\bx_1,\bx_2;\omega).
\end{equation}
The ground-state total energy can also be extracted from $G$ using the Galitskii-Migdal (GM) formula \citep{Galitskii_1958}
\begin{equation}
\label{Eqn:GM}
	E_0^\text{GM} = - \frac{\I}{2}\int \dbx_1 \lim_{2 \to 1^+}\qty[ \I \pdv{}{t_1} + h(\br_1) ] G(1,2),
\end{equation}
where $1 \equiv (\bx_1,t_1)$ is a space-spin plus time composite variable and $h(\br{})=-\nabla/2+v_\text{ext}(\br{})$ is the one-body Hamiltonian, $v_\text{ext}(\br{})$ being the local external potential. 

The one-body Green's function can be obtained by solving a Dyson equation of the form $G=G_0+G_0\Sigma G$, where $G_0$ is the non-interacting Green's function and the self-energy $\Sigma$ is an effective potential which contains all the many-body effects of the system under study. In practice, $\Sigma$ must be approximated and a well-known approximation is the so-called $GW$ approximation in which the self-energy reads $\Sigma^{GW}=v_H + \I GW$, where $v_H$ is the classical Hartree potential, and $W=\veps^{-1}v_c$ is the dynamically screened Coulomb interaction, with $\veps^{-1}$ the inverse dielectric function and $v_c$ the bare Coulomb interaction. \citep{Hedin_1965}

The equations stemming from the $GW$ approximation should, in principle, be solved self-consistently, since $\Sigma$ is a functional of $G$. \citep{Hedin_1965} Self-consistency, however, is computationally demanding, and one often performs a single $GW$ correction (for example using $G_0$ as starting point one builds $W$ and $\Sigma^{GW}$ as $\Sigma^{GW}=v_{\text{H}}+\I G_0W_0$, with $v_{\text{H}}=-\I v_cG_0$ and $W_0=[1+\I v_cG_0G_0]^{-1}v_c$, from which  $G=\{1-G_0\Sigma^{GW}[G_0]\}^{-1} G_0$). This cost-saving and popular strategy is known as one-shot $GW$. The main drawback of the one-shot $GW$ method is its dependence on the starting point (\ie, the orbitals and energies of the HF or KS mean-field eigenstates) originating from its perturbative nature. To overcome this problem, one can introduce some level of self-consistency. Removal/addition energies are thus obtained by solving iteratively the so-called quasiparticle equation
\begin{equation}
\label{Eqn:QP}
	\omega = \eps_i^\text{HF} + \mel{\MO{i}^\text{HF}}{\Sigma_\text{c}^{GW}(\omega)}{\MO{i}^\text{HF}}.
\end{equation}
Here, we choose to start from HF spatial orbitals $\MO{i}^\text{HF}(\br)$ and energies $\eps_i^{\text{HF}}$, which are corrected by the (real part of the) correlation contribution of the $GW$ self-energy $\Sigma^{GW}_\text{c}=\Sigma^{GW}-\Sigma_\text{HF}$, where $\Sigma_\text{HF}=v_H+\I v_c G$ is the HF (hartree plus exchange) contribution to the self-energy. $\Sigma^{GW}_c$ is evaluated with $G_{\text{HF}}$ at the first iteration, where $G_{\text{HF}}$ is the self-consistent solution of $G_{\text{HF}}=G_0+G_0\Sigma^{\text{HF}}G_{\text{HF}}$. At the $n$-th iteration, $\Sigma^{GW}_c$ is evaluated as $\Sigma^{GW}_c[G^{n-1}]$, where $G^{n-1}$ has poles at the energies from the $(n-1)$-th iteration of Eq.~\eqref{Eqn:QP} and corresponding weights obtained from the $Z$ factors given in Eq.~\eqref{Eqn:Z}.
As a non-linear equation, Eq.~\eqref{Eqn:QP} has potentially many solutions $\eps_{i,\nu}^{GW}$.
The  so-called quasiparticle (QP) solution $\eps_{i,\nu=0}^{GW} \equiv \eps_{i}^\text{QP}$ has the largest  renormalization factor (or spectral intensity)
\begin{equation}
\label{Eqn:Z}
	Z_{i,\nu} = \qty[ 1 - \left. \mel{\MO{i}^\text{HF}}{\pdv{\Sigma_{\text{c}}^{GW}(\omega)}{\omega}}{\MO{i}^\text{HF}} \right|_{\omega = \eps_{i,\nu}^{GW}} ]^{-1},
\end{equation}
while the satellite (sat) peaks $\eps_{i,\nu>0}^{GW} \equiv \eps_{i,\nu}^\text{sat}$ share the remaining of the spectral weight.
Moreover, one can show that the following sum rule is fulfilled \citep{vonBarth_1996}
\begin{equation}
\label{Eqn:Zsum}
	\sum_{\nu} Z_{i,\nu} = 1,
\end{equation}
where the sum runs over all the solutions of the quasiparticle equation for a given mean-field eigenstate $i$.
Throughout this article, $i$, $j$, $k$, and $l$ denote general spatial orbitals, $a$ and $b$ refer to occupied orbitals, $r$ and $s$ to unoccupied orbitals, while $m$ labels single excitations $a \to r$.

In eigenvalue self-consistent $GW$ (commonly abbreviated as ev$GW$), \citep{Hybertsen_1986,Shishkin_2007,Blase_2011,Faber_2011,Rangel_2016,Gui_2018} one only updates the poles of $G$, while keeping fix the orbitals (or weights). $G$ is then used to build $\Sigma^{GW}$ and $W$. At the $n$th iteration, the removal/addition energies are obtained from the $GW$ quasiparticle solutions computed from $G_{n-1}W(G_{n-1})$ where the satellites are discarded at each iteration. Nonetheless, at the final iteration one can keep the satellite energies to get the full spectral function [see Eq.~\eqref{Eqn:spectral_functions}]. In fully self-consistent $GW$ (sc$GW$),  \citep{Caruso_2012,Caruso_2013, Caruso_2013a,Caruso_2013b,Koval_2014} one updates the poles and weights of $G$ retaining quasiparticle and satellite energies at each iteration. 

It is instructive to mention that, for a conserving approximation, the sum of the intensities corresponding to removal energies equals the number of electrons, \ie, $\sum_{\eps_{i,\nu}^{GW}<\mu} Z_{i,\nu}=\Nel$. sc$GW$ is an example of conserving approximations, while, in general, the one-shot $GW$ does not conserve the number of electrons.

\subsection{Bethe-Salpeter equation}

\subsubsection{Neutral excitations}
Linear response theory \citep{Oddershede_1977,Casida_1995,Petersilka_1996} in MBPT is described by the Bethe-Salpeter equation. \citep{Strinati_1988} 
The standard BSE within the static $GW$ approximation (referred to as BSE@$GW$ in this work, which means the use of $GW$ quasiparticle energies to build the independent-particle excitation energies and of the $GW$ self-energy to build the static exchange-correlation kernel) can be recast, assuming a closed-shell reference state, as a non-Hermitian eigenvalue problem known as Casida equations:
\begin{equation}
	\begin{pmatrix}
		\bA{\IS}	&	\bB{\IS}	\\
		-\bB{\IS}	&	-\bA{\IS}	\\
	\end{pmatrix}
	\begin{pmatrix}
		\bX{\IS}_m	\\
		\bY{\IS}_m	\\
	\end{pmatrix}
	=
	\Om{m}{\IS}
	\begin{pmatrix}
		\bX{\IS}_m	\\
		\bY{\IS}_m	\\
	\end{pmatrix},
	\label{Eqn:EIGENVALUE_PB}
\end{equation}
where $\Om{m}{\IS}$ is the $m$th excitation energy with eigenvector $\T{(\bX{\IS}_m \, \bY{\IS}_m)}$ at interaction strength $\IS$, $\T{}$ is the matrix transpose, and we have assumed real-valued spatial orbitals. The non-interacting and physical systems correspond to $\IS = 0$ and $1$, respectively. 
The matrices $\bA{\IS}$ and $\bB{\IS}$ are of size $\Occ \Vir \times \Occ \Vir$, where $\Occ$ and $\Vir$ are the number of occupied and virtual orbitals, respectively, and $\Occ + \Vir$ is the total number of spatial orbitals.
Introducing the so-called Mulliken notation for the bare two-electron integrals
\begin{equation}
	\ERI{ij}{kl} = \iint \dbr_1 \dbr_2 \MO{i}(\br_1) \MO{j}(\br_1) v_c(\br_1-\br_2)\MO{k}(\br_2) \MO{l}(\br_2),
\end{equation}
and the corresponding (static) screened Coulomb potential matrix elements 
\begin{equation}
	W_{ij,kl}(\omega=0) = \iint \dbr_1 \dbr_2  \MO{i}(\br_1) \MO{j}(\br_1) W(\br_1,\br_2; \omega=0 ) \MO{k}(\br_2) \MO{l}(\br_2),
\end{equation}
the BSE matrix elements read \citep{Maggio_2016}
\begin{subequations}
\begin{align}
	\label{eq:LR_BSE-A}
	\A{ar,bs}{\IS,\sigma\sigma'} & = \delta_{ab} \delta_{rs} (\eps_r^\text{QP} - \eps_a^\text{QP}) + \IS \qty[ \alpha_{\sigma\sigma'} \ERI{ar}{sb}   -  W_{ab,sr}(\omega=0) ],
	\\ 
	\label{eq:LR_BSE-B}
	\B{ar,bs}{\IS,\sigma\sigma'} & =  \IS \qty[ \alpha_{\sigma\sigma'} \ERI{ar}{bs} -   W_{as,br}(\omega=0) ],
\end{align}
\end{subequations}
where $\eps_i^\text{QP}$ are the $GW$ quasiparticle energies, and $\alpha_{\singlet}=2$ and $\alpha_{\triplet}=0$ for singlet (\ie, spin-conserved) and triplet (\ie, spin-flip) excitations, respectively. 

In the absence of instabilities (\ie, when $\bA{\IS} - \bB{\IS}$ is positive-definite), \citep{Dreuw_2005} Eq.~\eqref{Eqn:EIGENVALUE_PB} is usually transformed into an Hermitian eigenvalue problem of half the dimension
\begin{equation}
	(\bA{\IS} - \bB{\IS})^{1/2} (\bA{\IS} + \bB{\IS}) (\bA{\IS} - \bB{\IS})^{1/2} \bZ{m}{\IS} = (\Om{m}{\IS})^2 \bZ{m}{\IS},
	\label{Eqn:BSE_smaller}
\end{equation}
where the excitation amplitudes are
\begin{subequations}
\begin{align}
	(\bX{\IS} + \bY{\IS})_m = (\Om{m}{\IS})^{-1/2} (\bA{\IS} - \bB{\IS})^{+1/2} \bZ{m}{\IS},
	\\
	(\bX{\IS} - \bY{\IS})_m = (\Om{m}{\IS})^{+1/2} (\bA{\IS} - \bB{\IS})^{-1/2} \bZ{m}{\IS}.
\end{align}
\end{subequations}
Singlet ($\Om{m}{\singlet} \equiv \Om{m}{\IS=1,\singlet}$) and triplet ($\Om{m}{\triplet} \equiv \Om{m}{\IS=1,\triplet}$) excitation energies are obtained by diagonalizing Eq.~\eqref{Eqn:EIGENVALUE_PB} at $\IS=1$.

\subsubsection{Correlation energies}
Our goal here is to compare the BSE correlation energy $\EcBSE$ obtained using two formulas, namely the trace (or plasmon) formula \citep{Ring-Schuck,Rowe_1968} 
and the expression obtained using the adiabatic-connection fluctuation-dissipation theorem (ACFDT) formalism. \citep{Furche_2005,Toulouse_2009,Toulouse_2010,Hellgren_JCP2010,Hesselmann_PRL2011,Angyan_2011,Colonna_2014,Maggio_2016,Holzer_2018b,Loos_2020e} The two approaches have been recently compared at the random-phase approximation (RPA) level for the case of \ce{Be2}, \citep{Li_2020} showing similar improved performances at the RPA@$GW$@PBE level with respect to the RPA@PBE level and an impressive accuracy by introducing BSE (BSE@$GW$@HF) correction in the trace formula. Here we would like to get more insights into the quality of these two approaches.

The ground-state correlation energy within the trace formula is calculated as
\begin{equation}
\begin{split}
 	\label{Eqn:tracef}
	\EcTrBSE 
	& = \EcTr{\singlet} + \EcTr{\triplet} 
	\\
	& = \frac{1}{2}\qty[ \sum_{m} \Om{m}{\singlet} - \Tr(\bA{\singlet}) ] + \frac{1}{2}\qty[ \sum_{m} \Om{m}{\triplet} - \Tr(\bA{\triplet}) ],
\end{split}
\end{equation}
where $\bA{\sigma\sigma'} \equiv \bA{\IS=1,\sigma\sigma'}$ is defined in Eq.~\eqref{eq:LR_BSE-A}, and $\Tr$ denotes the matrix trace. We note that the trace formula is an approximate expression of the correlation energy since it relies on the so-called quasi-boson approximation
and on the killing condition on the zeroth-order Slater determinant ground state (see Ref.~\citep{Li_2020} for more details).
Note that here both sums in Eq.~\eqref{Eqn:tracef} run over all resonant (hence real- and complex-valued) excitation energies while they are usually restricted to the real-valued resonant BSE excitation energies. Thus, the Tr@BSE correlation energy is potentially a complex-valued function in the presence of singlet and/or triplet instabilities.

The ACFDT formalism, \citep{Furche_2005} instead, provides an in-principle exact expression for the correlation energy within time-dependent density-functional theory (TDDFT). \citep{Runge_1984,Petersilka_1996,UlrichBook} In practice, however, one always ends up with an approximate expression, which quality relies on the approximations to the exchange-correlation potential of the KS system and to the kernel of the TDDFT linear response equations. In this work, therefore, we use the ACFDT expression within the BSE formalism and we explore how well it performs and how it compares to the trace formula \eqref{Eqn:tracef}.

Within the ACFDT framework, only the singlet states do contribute for a closed-shell ground state, and the ground-state BSE correlation energy 
\begin{equation}
	\EcACBSE = \frac{1}{2} \int_0^1 d\IS \Tr(\bK^{\singlet} \bP{\IS,\singlet}) 
	\label{eqn:intAC}
\end{equation}
is obtained via integration along the adiabatic connection path from the non-interacting system at $\IS = 0$ to the physical system $\IS = 1$, where
\begin{equation}
\label{eq:K}
	\bK = 
	\begin{pmatrix}
		\btA{\IS=1}	&	\bB{\IS=1}	\\
		\bB{\IS=1}	&	\btA{\IS=1}	\\
	\end{pmatrix}
\end{equation}
is the interaction kernel, \citep{Angyan_2011,Holzer_2018b,Loos_2020e} $\tA{ar,bs}{\IS,\sigma\sigma'} = \alpha_{\sigma\sigma'} \IS \ERI{ar}{sb}$, and
\begin{equation}
\label{eq:2DM}
	\bP{\IS} = 
	\begin{pmatrix}
		\bY{\IS} \T{(\bY{\IS})}		&	\bY{\IS} \T{(\bX{\IS})}	\\
		\bX{\IS} \T{(\bY{\IS})}		&	\bX{\IS} \T{(\bX{\IS})}	\\
	\end{pmatrix}
	-
	\begin{pmatrix}
		\bO		&	\bO	\\
		\bO		&	\bI	\\
	\end{pmatrix}
\end{equation} 
is the correlation part of the two-body density matrix at interaction strength $\IS$.
Here again, the AC@BSE correlation energy might become complex-valued in the presence of singlet instabilities.

Note that the trace and ACFDT formulas yield, for any set of eigenstates, the same correlation energy at the RPA level. \citep{Angyan_2011}
Moreover, in contrast to density-functional theory where the electron density is fixed along the adiabatic path, \citep{Langreth_1975,Gunnarsson_1976,Zhang_2004} at the BSE@$GW$ level, the density is not maintained as $\IS$ varies. Therefore, an additional contribution to Eq.~\eqref{eqn:intAC} originating from the variation of the Green's function along the adiabatic connection should, in principle, be added. However, as commonly done within RPA  \citep{Toulouse_2009,Toulouse_2010,Angyan_2011,Colonna_2014} and BSE, \citep{Holzer_2018b,Loos_2020e} we neglect this additional contribution.

\begin{figure*}
  \includegraphics[width=1.0\textwidth]{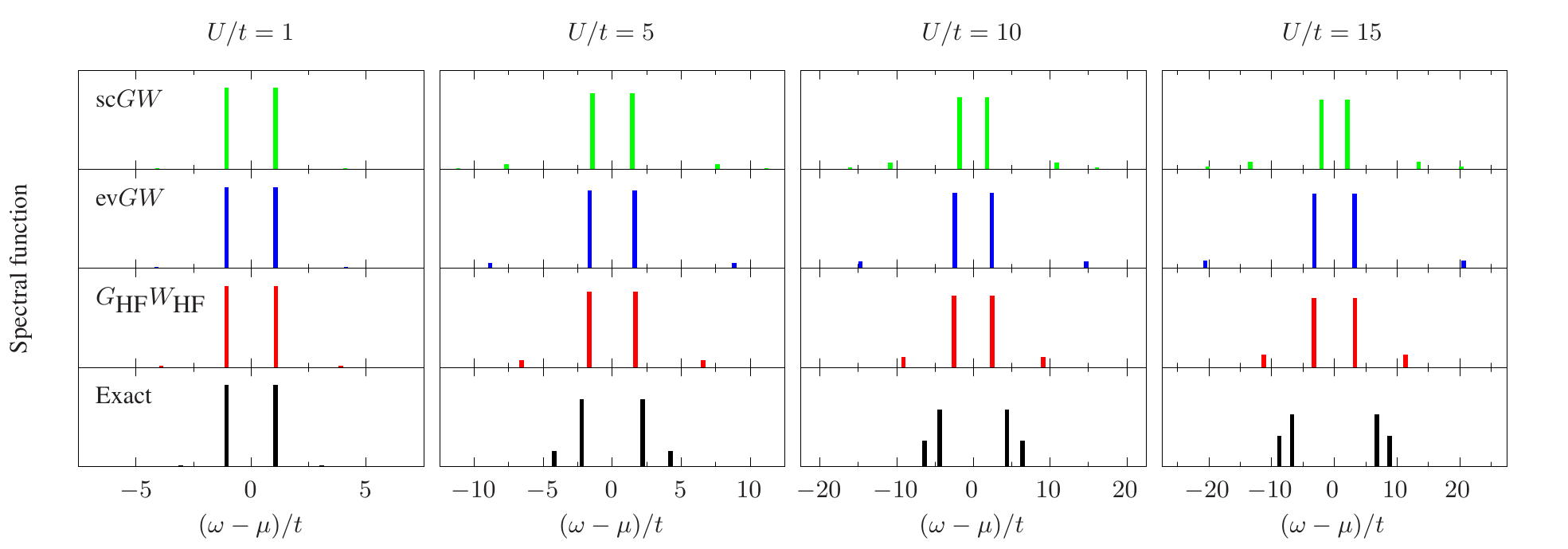}
 \caption{Spectral function of $G$ [see Eq.~\eqref{Eqn:spectral_functions}] as a function of $(\omega - \mu)/t$ (where $\mu = U/2$ is the chemical potential) at various values of the ratio $U/t$ ($U/t=1$, $5$, $10$, and $15$) for different levels of theory: exact (black), $G_\text{HF}W_\text{HF}$ (red), ev$GW$ (blue), and sc$GW$ (green). All approximate schemes are obtained using $G_\text{HF}$ as starting point.
 }
\label{fig:spectrumGW@HF}
\end{figure*}

\begin{squeezetable}
\begin{table*}[t]
    \caption{Numerical values of quasiparticle energy $\eps_\an^\text{QP}$ and satellite energy $\eps_\an^{\text{sat}}$ (anti-bonding components) and respective intensities ($Z_\an^\text{QP}$ and $Z_\an^{\text{sat}}$) for the spectral functions presented in Fig.~\ref{fig:spectrumGW@HF}. Energies are relative to the chemical potential $\mu=U/2$. All spectral functions presented in Fig.~\ref{fig:spectrumGW@HF} are symmetric with respect to $\mu$, which means that $\eps_\bn^\text{QP/sat}=-\eps_\an^\text{QP/sat}$ and $Z_\bn^\text{QP/sat}=Z_\an^\text{QP/sat}.$}
    \label{tab:GW}
    \begin{ruledtabular}
    \begin{tabular}{ccccccccccccccccc}
     & \multicolumn{4}{c}{$\eps_\an^\text{QP}$}
    & \multicolumn{4}{c}{$\eps_\an^{\text{sat}}$}
    & \multicolumn{4}{c}{$Z_\an^\text{QP}$}
    & \multicolumn{4}{c}{$Z_\an^{\text{sat}}$}\\
    \cline{2-5} \cline{6-9} \cline{10-13} \cline{14-17}
    $U/t$ & exact & $G_\text{HF}W_\text{HF}$ & ev$GW$ & sc$GW$ & exact & $G_\text{HF}W_\text{HF}$ & ev$GW$ & sc$GW$ & exact & $G_\text{HF}W_\text{HF}$ & ev$GW$ & sc$GW$ & exact & $G_\text{HF}W_\text{HF}$ & ev$GW$ & sc$GW$\\
    \hline
1 & 1.0615 & 1.0721 & 1.0702 & 1.0651 & 3.0615 & 3.9006 & 4.1175 & 4.0793 & 0.9851 & 0.9855 & 0.9864 & 0.9861 & 0.0149 & 0.0145 & 0.0135 & 0.0132 \\
5 & 2.2016 & 1.6739 & 1.6302 & 1.4334 & 4.2016 & 6.5728 & 8.8364 & 7.6389 & 0.8123 & 0.9183 & 0.9398 & 0.9239 & 0.1876 & 0.0817 & 0.0602 & 0.0593 \\
10 & 4.3852 & 2.4893 & 2.4001 & 1.7787 & 6.3852 & 9.1225 & 14.7136 & 10.8296 & 0.6857 & 0.8717 & 0.9182 & 0.8777 & 0.3143 & 0.1282 & 0.0818 & 0.0823 \\
15 & 6.7621 & 3.2887 & 3.1813 & 2.0542 & 8.7621 & 11.2887 & 20.5769 & 13.3847 & 0.6288 & 0.8430 & 0.9082 & 0.8472 & 0.3712 & 0.1570 & 0.0918 & 0.0934\\
    \end{tabular}
    \end{ruledtabular}
\end{table*}
\end{squeezetable}

\section{Results}
\label{sec:Results}
As discussed in Sec.~\ref{sec:intro}, in this work, we consider the (symmetric) Hubbard dimer as test case, which is governed by the following Hamiltonian
\begin{equation}
 \label{Eqn:Hamiltonian}
 \Hat{H} 
 =-t \sum_{\sigma=\uparrow,\downarrow} \qty( \Hat{c}^{\dagger}_{1\sigma} \Hat{c}_{2\sigma} + \Hat{c}^{\dagger}_{2\sigma} \Hat{c}_{1\sigma} )
 + U \qty( \Hat{n}_{1\uparrow} \Hat{n}_{1\downarrow}+\Hat{n}_{2\uparrow} \Hat{n}_{2\downarrow} ).
 \end{equation}
Here $\Hat{n}_{1\sigma}=\Hat{c}^{\dagger}_{1\sigma} \Hat{c}_{1\sigma}$ ($\Hat{n}_{2\sigma}=\Hat{c}^{\dagger}_{2\sigma} \Hat{c}_{2\sigma}$) is the spin density operator on site 1 (site 2), $\Hat{c}^{\dagger}_{1\sigma}$ and $\Hat{c}_{1\sigma}$ ($\Hat{c}^{\dagger}_{2\sigma}$ and $\Hat{c}_{2\sigma}$) are the creation and annihilation operators for an electron at site 1 (site 2) with spin $\sigma$, $U$ is the on-site (spin-independent) interaction, and $-t$ is the hopping kinetic energy. The physics of the Hubbard model arises from the competition between the hopping term, which prefers to delocalize electrons, and the on-site interaction, which favors localization. The ratio $U/t$ is a measure for the relative contribution of both terms and is the intrinsic, dimensionless coupling constant of the Hubbard model, which we use in the following. 
In this work we consider the dimer at one-half filling.

\subsection{Quasiparticle energies in the $GW$ approximation}
We test different flavors of self-consistency in $GW$ calculations: one-shot $GW$, ev$GW$, partial self-consistency through the alignment of the chemical potential (psc$GW$), where we shift $G_0$ or $G_\text{HF}$ in such a way that the resulting $G$ has the same chemical potential than the shifted $G_0$ or shifted $G_\text{HF}$,\citep{Schindlmayr_1997} and sc$GW$. 
In the one-shot formalism, we also test two different starting points: the truly non-interacting Green's function $G_0$ ($U=0$) and the HF Green's function $G_\text{HF}$.
These two schemes are respectively labeled as $G_0W_0$ and $G_\text{HF}W_\text{HF}$ in the following. 

The $G_0W_0$ self-energy (in the site basis) and removal/addition energies are already given in Ref.~\citep{Romaniello_2012} for the Hubbard dimer at one-half filling. For completeness we report them in Appendix \ref{app:G0W0}, together with the renormalization factors, which are discussed in Sec.~\ref{G0_SP}.

Starting from $G_\text{HF}$, which reads
\begin{equation}
\label{Eqn:G_site}
    G_{\text{HF},IJ}(\omega) = \frac{1}{2}
    \qty[ \frac{(-1)^{(I-J)}}{\omega-(t+U/2)+\I\eta}
    +\frac{1}{\omega+(t-U/2)-\I\eta} ],
\end{equation}
where $I$ and $J$ run over the sites, the (correlation part of the) $G_\text{HF}W_\text{HF}$ self-energy is $\Sigma^{GW}_{\text{c},IJ}(\omega) = \Sigma^{GW}_{IJ}(\omega)-\delta_{IJ} U/2$ with
\begin{equation}
\label{Eqn:Sig_site}
    \Sigma^{GW}_{\text{c},IJ}(\omega) =\frac{U^2t}{2h}
    \left[ \frac{1}{\omega-(t+h+U/2)+\I\eta}  +\frac{(-1)^{I-J}}{\omega+(t+h-U/2)-\I\eta}\right],
\end{equation}
where $h=\sqrt{4t^2+4Ut}$. Here we used the following expression for the polarizability $P=-iGG$ with elements
\begin{equation}
\label{Eqn:P_site}
    P_{IJ}(\omega)=\frac{(-1)^{I-J}}{4}
    \qty[\frac{1}{\omega-2t+\I\eta}-\frac{1}{\omega+2t-\I\eta}]
\end{equation}
to build the screened interaction $W=v_c +v_cPW$, whose only non-zero matrix elements reads 
\begin{equation}
\label{Eqn:W_site}
    W_{II,JJ}(\omega)
    =U\delta_{IJ}+(-1)^{I-J}\frac{U^2t}{h}
    \qty[\frac{1}{\omega-h+\I\eta}-\frac{1}{\omega+h-\I\eta}]
\end{equation}
due to the local nature of the electron-electron interaction.
The quantities defined in Eqs.~\eqref{Eqn:G_site}, \eqref{Eqn:Sig_site}, \eqref{Eqn:P_site}, and \eqref{Eqn:W_site} can then be transformed to the bonding ($\bn$) and antibonding ($\an$) basis [which is used to recast the BSE as Eq.~\eqref{Eqn:EIGENVALUE_PB}] thanks to the following expressions:
\begin{align}
    \ket{\bn} & = \frac{\ket{1} + \ket{2}}{\sqrt{2}},
    &
    \ket{\an} & = \frac{\ket{1} - \ket{2}}{\sqrt{2}}.
\end{align}
Therefore, the one-shot removal/addition energies read
\begin{subequations}
\begin{align}
	\eps_{1,\pm} & = + \frac{h}{2} +  \frac{U}{2} \pm \frac{\sqrt{
	(h+2t)^2 + 4 t U^2/h}}{2},
	\label{Eqn:QP_b}
	\\
	\eps_{2,\pm} & =   -\frac{h}{2} +  \frac{U}{2} \pm \frac{\sqrt{
	(h+2t)^2+ 4 t U^2/h}}{2},
	\label{Eqn:QP_a}
\end{align}
\end{subequations}
with the quasiparticle solutions being $\eps_\bn^\text{QP}=\eps_{1,-}$ and $\eps_\an^\text{QP}=\eps_{2,+}$, which correspond to the bonding and antibonding energies, respectively.
As readily seen in Eqs.~\eqref{Eqn:QP_b} and \eqref{Eqn:QP_a}, in addition to the quasiparticle, there is a unique satellite per eigenstate given by $\eps_\bn^\text{sat}=\eps_{1,+}$ and $\eps_\an^\text{sat}=\eps_{2,-}$.
Moreover, the closed-form expression of the renormalization factors [see Eq.~\eqref{Eqn:Z}] read
\begin{equation}
\label{Eqn:Z_QP}
Z^{\text{QP}}_{\bn/\an}=\frac{t \qty[h^2 + 2 h t + 2 U^2 + h \sqrt{  (h + 2 t)^2 + 4 t U^2/h}]}{h^3 + 
 4 h^2 t + 4 h t^2 + 4 t U^2 - h^2 \sqrt{  (h + 2 t)^2 + 4 t U^2/h}  }  
\end{equation}
and $Z^{\text{sat}}_{\bn/\an}=1-Z^{\text{QP}}_{\bn/\an}$.

The ev$GW$ and sc$GW$ calculations were performed numerically using the meromorphic representation of $G$, following Ref.~\citep{vonFriesen_2010} with some slight modifications (see Appendix \ref{app:meromorphic} for more details).
At each iteration, the solution of the Dyson equations for $G$ and $W$ (see Sec.~\ref{sec:GWA}) produces extra poles. 
In order to keep the number of poles under control in sc$GW$, the poles with intensities smaller than a user-defined threshold (set from $10^{-4}$ to $10^{-6}$ depending on the ratio $U/t$) are discarded and the corresponding spectral weight is redistributed among the remaining poles.

\begin{figure*}
  \includegraphics[width=0.44\textwidth]{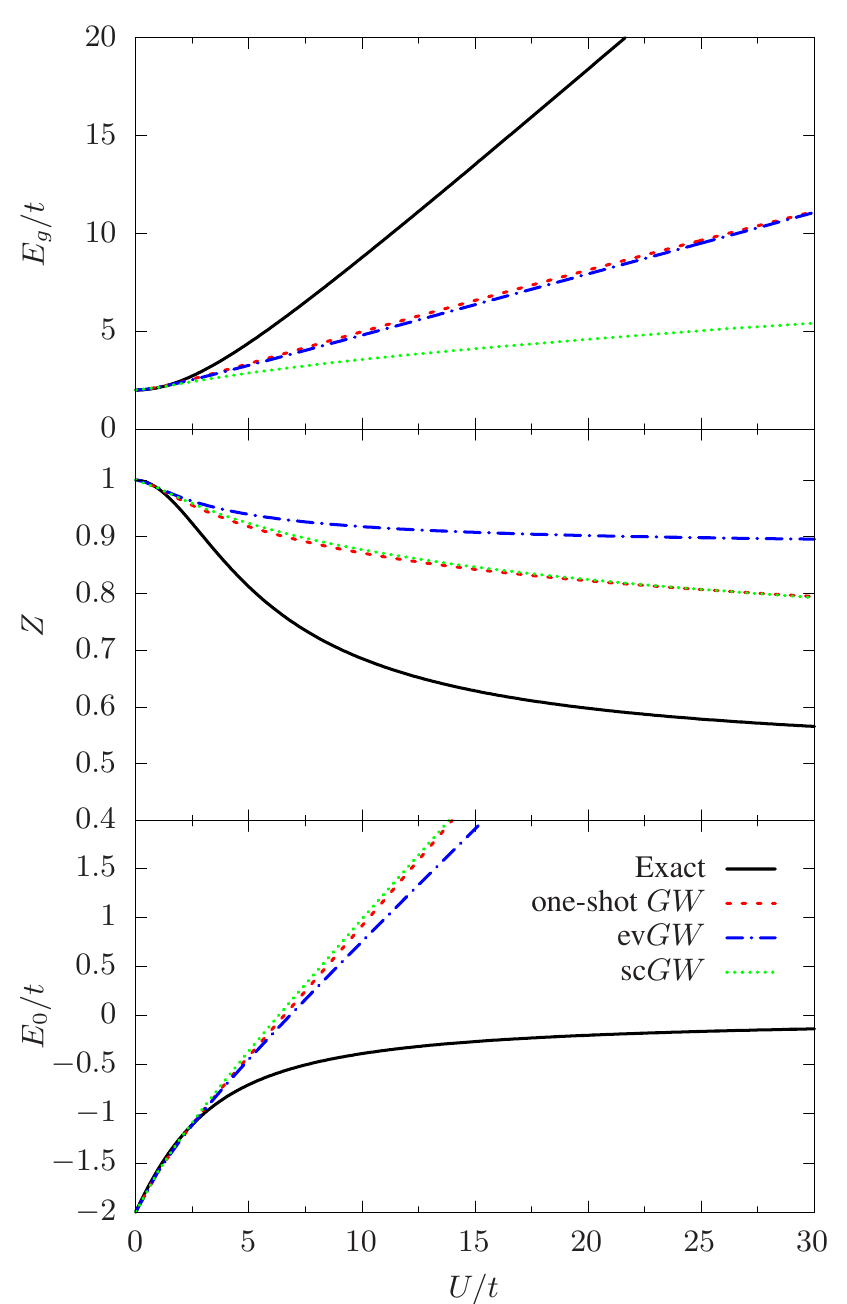}
  \hspace{0.05\textwidth}
  \includegraphics[width=0.44\textwidth]{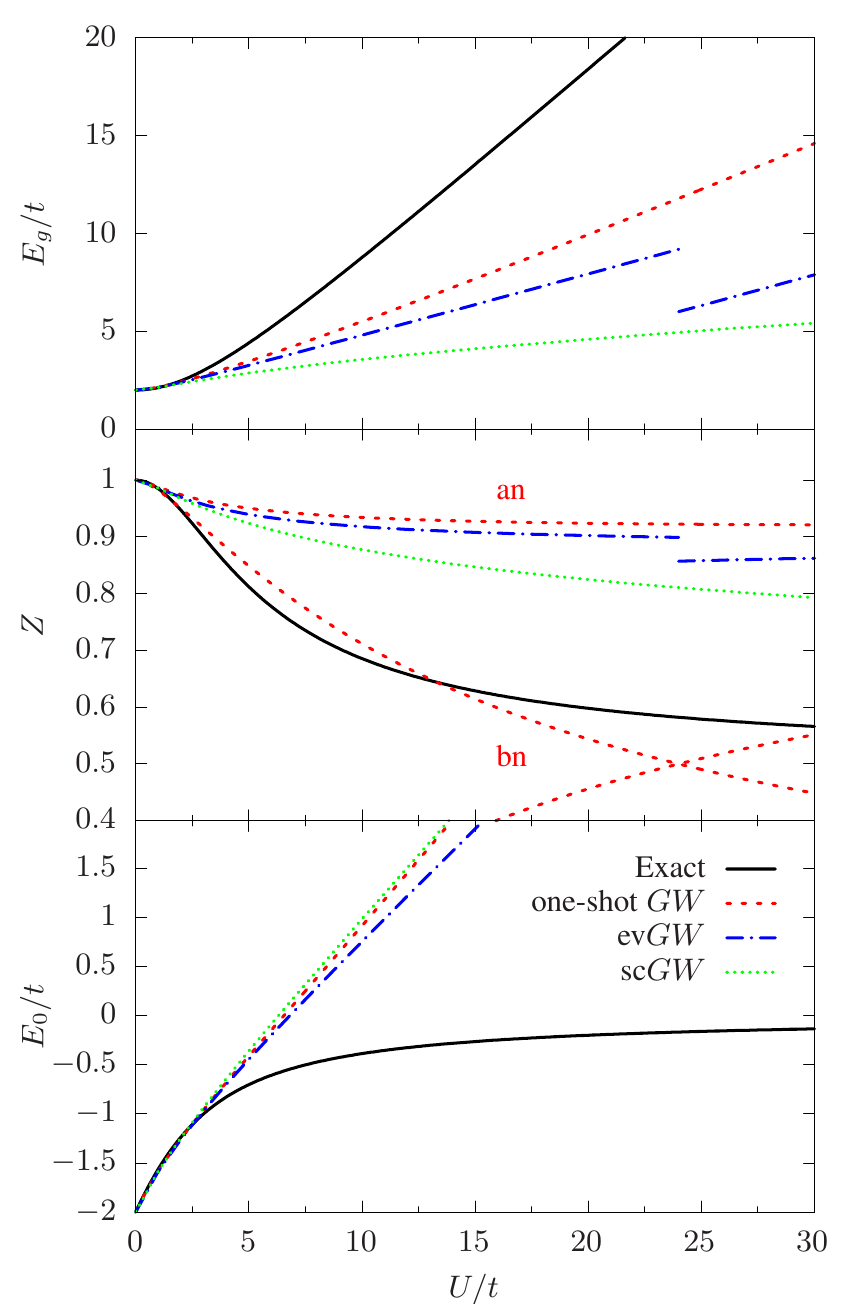}
 \caption{Fundamental gap ($\Eg$), quasiparticle weight factors ($Z_{\bn/\an}^\text{QP}$), and ground state energy ($E_0$) as functions of $U/t$ obtained from one-shot $GW$ (dashed red line), ev$GW$ (dashed-dotted blue line), sc$GW$ (dotted green line) using $G_{\text{HF}}$ (left) or $G_0$ (right) as starting point. The black curves are the exact results.}
\label{fig:bandgapGW}
\end{figure*}

In Fig.~\ref{fig:spectrumGW@HF}, we present the spectral function of $G$ [see Eq.~\eqref{Eqn:spectral_functions}] for different values of the ratio $U/t$ ($U/t=1$, $5$, $10$, and $15$) and using $G_{\text{HF}}$ as starting point. We consider three $GW$ variants: $G_{\text{HF}}W_{\text{HF}}$, ev$GW$, and sc$GW$.
For $U/t\lesssim3$, all the schemes considered here provide a faithful description of the quasiparticle energies. For larger $U/t$, $GW$ (regardless of the level of self-consistency) tends to underestimate the fundamental gap $\Eg$ [see Eq.~\eqref{Eqn:EgFun}], as shown in the upper left panel of Fig.~\ref{fig:bandgapGW}. $G_{\text{HF}}W_{\text{HF}}$ and ev$GW$ give a very similar estimate of $\Eg$, whereas the quasiparticle intensity $Z_{\bn/\an}^\text{QP}$ defined in Eq.~\eqref{Eqn:Z_QP}
is quite different and overestimated by both methods, at least in the range of $U/t$ considered in Fig.~\ref{fig:bandgapGW} (center left panel).

The main effects of full self-consistency are the reduction of $\Eg$ (see upper left panel of Fig.~\ref{fig:bandgapGW}), and the creation of extra satellites with decreasing intensity (see upper panel of Fig.~\ref{fig:spectrumGW@HF}).
For small $U/t$, the fundamental gap is similar to the one predicted by other methods while for increasing $U/t$ the agreement worsen and $\Eg$ is grossly underestimated. The quasiparticle intensity is very similar to the one predicted by $G_{\text{HF}}W_{\text{HF}}$.
Concerning the position of the satellites, we observe that the one-shot $G_{\text{HF}}W_{\text{HF}}$ scheme gives the most promising results. Numerical values of quasiparticle and first satellite energies as well as their respective intensities in the spectral functions presented in Fig.~\ref{fig:spectrumGW@HF} are gathered in Table \ref{tab:GW}. 

We notice that a similar analysis for $H_2$ in a minimal basis has been presented in Ref.~\citep{Hellgren_2015} with analogous conclusions.

For the sake of completeness, we also report in the bottom left panel of Fig.~\ref{fig:bandgapGW} the total energy calculated using the Galinski-Migdal formula [see Eq.~\eqref{Eqn:GM}]. Since the Galinski-Migdal total energy is not stationary with respect to changes in $G$, one gets meaningful energies only at self-consistency. However, for the Hubbard dimer, we do not observe a significant impact of self-consistency, as one can see from Fig.~\ref{fig:spectrumGW@HF} by comparing the total energy at the $G_{\text{HF}}W_{\text{HF}}$, ev$GW$, and sc$GW$ levels. For each of these schemes which correspond to a different level of self-consistency, the Galinski-Migdal formula provides accurate total energies only for relatively small $U/t$ ($\lesssim 3$).

If we consider $G_{\text{HF}}$ as starting point and we define the chemical potential as $\mu=(\eps_\an^\text{QP}+\eps_\bn^\text{QP})/2$, then the alignment of the chemical potential has no effect on the spectrum, this means that $G_{\text{HF}}W_{\text{HF}}$ and psc$GW$ are equivalent.

\subsubsection{$G_0$: a bad starting point\label{G0_SP}}
In the following we will illustrate how the starting point can influence the resulting quasiparticle energies.
The Green's function obtained from the one-shot $G_0W_0$ does not satisfy particle-hole symmetry, the fundamental gap is underestimated  (top right panel of Fig.~\ref{fig:bandgapGW}) yet more accurate than $G_\text{HF}W_\text{HF}$ (top left panel of Fig.~\ref{fig:bandgapGW}), the quasiparticle intensity relative to the bonding component is close to the exact result up to $U/t\approx16$ (center right panel of Fig.~\ref{fig:bandgapGW}), while overestimated for the antibonding components. 
Moreover, we note that the intensities of the two poles of the bonding component crosses at $U/t=24$.
This means that if we sort the quasiparticle and the satellite according to their intensity at a given $U/t$, the nature of the two poles is interchanged when one increases $U/t$, which results in a discontinuity in the QP energy.
Meanwhile, the total number of particle is not conserved ($\Nel<2$). For $G_0W_0$ we found a small deviation from $\Nel=2$ for small $U/t$ (e.g. $\Nel=1.98828$ at $U=1$), which becomes larger by increasing the interaction (e.g. $\Nel=1.55485$ for $U/t=10$). Instead, starting from $G_\text{HF}$ the particle number is always conserved. We checked that for the self-consistent calculations the total particle number is conserved, as it should.

Considering $G_0$ as starting point in ev$GW$, we encounter the problem described in Ref.~\citep{Veril_2018}, namely the discontinuity of various key properties (such as the fundamental gap in the top right panel of Fig.~\ref{fig:bandgapGW}) with respect to the interaction strength $U/t$. 
This issue is solved, for the Hubbard dimer, by considering a better starting point or using the fully self-consistent scheme sc$GW$.
Note, however, that improving the starting point does not always cure the discontinuity problem as this issue stems from the quasiparticle approximation itself.
Full self-consistency, instead, avoids systematically discontinuities since no distinction is made between quasiparticle and satellites. Unfortunately, full self-consistency is much more involved from a computational point of view and, moreover, it does not give an overall improvement of the various properties of interest, at least for the Hubbard dimer, for which $G_{\text{HF}}W_{\text{HF}}$ is to be preferred. For more realistic (molecular) systems, it was shown in Ref.~\citep{Berger_2021} that the computationally cheaper self-consistent COHSEX scheme solves the problem of multiple quasiparticle solutions.

\subsection{BSE}
For the Hubbard dimer the matrices $\bA{\IS}$ and $\bB{\IS}$ in Eq.~\ref{Eqn:EIGENVALUE_PB} are just single matrix elements and they simply read, for both spin manifolds,
\begin{subequations}
\begin{align}
	\label{Eqn:AB_S}
	\A{}{\IS,\singlet} & =\Delta\eps^{GW}+ \IS \frac{U}{2},
	&
	\B{}{\IS,\singlet} & =  \IS \frac{U}{2} \qty(\frac{4 t U}{h^2} + 1 ),
	\\
	\label{Eqn:AB_T}
	\A{}{\IS,\triplet} & =\Delta\eps^{GW}- \IS \frac{U}{2},
	&
	\B{}{\IS,\triplet} & =  \IS \frac{U}{2} \qty(\frac{4 t U}{h^2} - 1 ),
\end{align}
\end{subequations}
while $\tA{}{\IS,\singlet} = \IS U$. We employ the screened Coulomb potential given in Eq.~\eqref{Eqn:W_site} at $\omega=0$ for the kernel,
and the $GW$ quasiparticle energies from Eqs.~\eqref{Eqn:QP_b} and \eqref{Eqn:QP_a} to build the $GW$ approximation of the fundamental gap $\Delta\eps^{GW} = \eps^{\text{QP}}_{\an}-\eps^{\text{QP}}_{\bn}$.
For comparison purposes, we also use the \textit{exact} quasiparticle energies [see Eq.~(C3) of Ref.~\citep{Romaniello_2012}], which consists in replacing $\Delta\eps^{GW}$ by the \textit{exact} fundamental gap $\Eg = \sqrt{16t^2+U^2}-2t$. 
In such a case, one is able to specifically test how accurate the BSE formalism is at catching the excitonic effect via the introduction of the screened Coulomb potential.

We notice that, within the so-called Tamm-Dancoff approximation (TDA) where one neglects the coupling matrix $\bB{\IS}$ between the resonant and anti-resonant parts of the BSE Hamiltonian [see Eq.~\eqref{Eqn:EIGENVALUE_PB}], BSE yields RPA with exchange (RPAx) excitation energies for the Hubbard dimer. This is the case also for approximations to the BSE kernel which are beyond $GW$, such as the T-matrix approximation. \citep{Romaniello_2012,Zhang_2017,Li_2021b}, and it is again related to the local nature of the electron-electron interaction. Hence, to test the effect of approximations on correlation for this model system we must go beyond the TDA.

\begin{figure*}
  \includegraphics[width=0.44\textwidth]{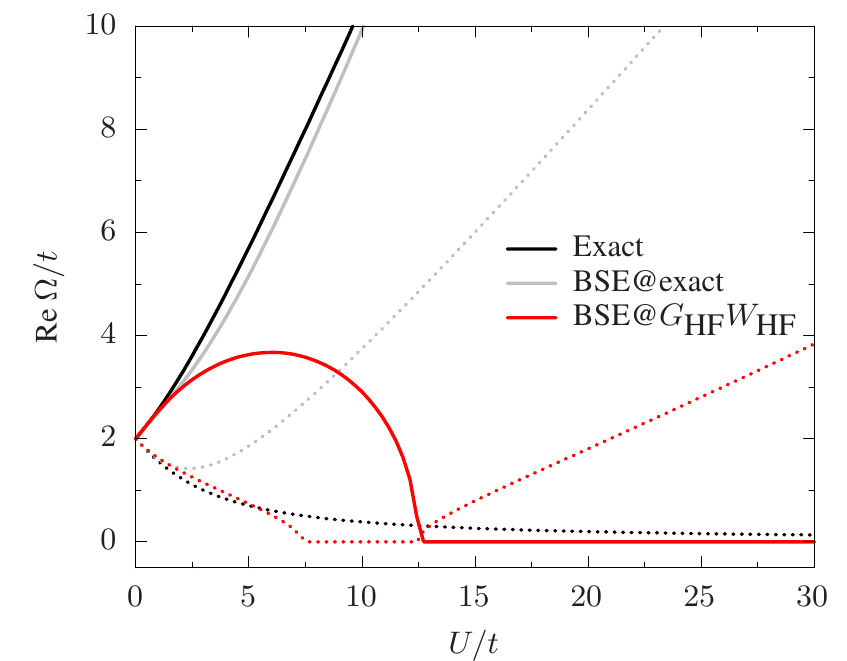}
  \includegraphics[width=0.44\textwidth]{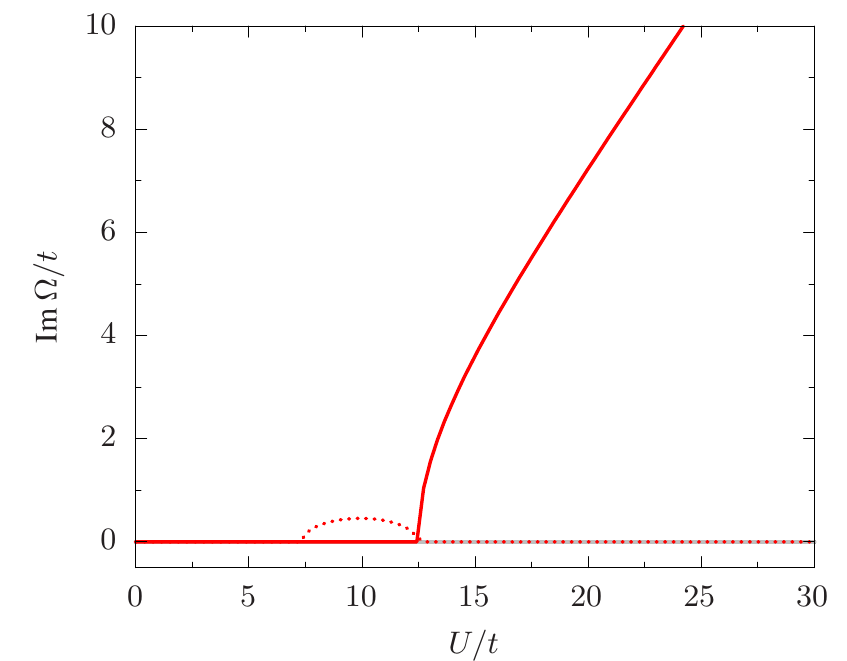}
 \caption{Real and imaginary parts of the singlet (solid) and triplet (dotted) neutral excitations, $\Om{1}{\singlet}$ and $\Om{1}{\triplet}$, as functions of $U/t$: exact (black), BSE with exact quasiparticle energies and $W_\text{HF}$ (gray), BSE@$G_\text{HF}W_\text{HF}$ (red).}
\label{fig:excittaions}
\end{figure*}

\subsubsection{Neutral excitations}
In Fig.~\ref{fig:excittaions}, we report the real part of the singlet and triplet excitation energies obtained from the solution of Eq.~\eqref{Eqn:EIGENVALUE_PB} for $\IS=1$. For comparison, we report also the exact excitation energies obtained as differences of the excited- and ground-state total energies of the Hubbard dimer obtained by diagonalizing the Hamiltonian \eqref{Eqn:Hamiltonian} in the Slater determinant basis $\{ \ket{1\uparrow,1\downarrow}, \ket{1\uparrow,2\downarrow},\ket{1\downarrow,2\uparrow},\ket{2\uparrow,2\downarrow}\}$ built from the sites (see Ref.~\citep{Romaniello_2009a} for the exact total energies).
For the singlet manifold, this yields, for the single excitation $\Om{1}{\singlet}$ and double excitation $\Om{2}{\singlet}$, the following expressions:
\begin{align}
	\label{eqn:exSinglet}
	\Om{1}{\singlet} & = \frac{1}{2}\qty(U+\sqrt{16 t^2 + U^2}), 
	&
	\Om{2}{\singlet} & = \sqrt{16 t^2 + U^2},
\end{align}
while the unique triplet transition energy is
\begin{equation}
	\label{eqn:exTriplet}
	\Om{1}{\triplet} = \frac{1}{2}\qty(-U+\sqrt{16 t^2 + U^2}).
\end{equation}
Of course, one cannot access the double excitation within the static approximation of BSE, \citep{Strinati_1988,Romaniello_2009b,Loos_2020h} so only the lowest singlet and triplet excitations, $\Om{1}{\singlet}$ and $\Om{1}{\triplet}$, are studied below.

Using one-shot $G_{\text{HF}}W_{\text{HF}}$ quasiparticle energies (BSE@$G_{\text{HF}}W_{\text{HF}}$) produces complex excitation energies (see right panel of Fig.~\ref{fig:excittaions}). We find the same scenario also with other flavors of $GW$ (not reported in the figure), such as sc$GW$.  The occurrence of complex poles and singlet/triplet instabilities at the BSE level are well documented \citep{Holzer_2018b,Blase_2020,Loos_2020e} and is not specific to the Hubbard dimer. For example, one finds complex poles also for \ce{H2} along its dissociation path, \citep{Li_2021a} but also for larger diatomic molecules. \citep{Loos_2020e}
For $U/t>12.4794$, the singlet energy becomes pure imaginary, the same is observed for the triplet energy for $7.3524<U/t<12.4794$. These two points corresponds to discontinuities in the first derivative of the excitation energies with respect to $U/t$ (see Fig.~\ref{fig:excittaions}).
The BSE excitation energies are good approximations to their exact analogs only for $U/t\lesssim 2$ for the singlet and $U/t\lesssim 6$ for the triplet.
Using exact quasiparticle energies instead produces real excitation energies, with the singlet energy in very good agreement with the exact result; the triplet energy, instead, largely overestimates the exact value. 
This seems to suggest that complex poles are caused by the approximate nature of the $GW$ quasiparticle energies, although, of course, the quality of the kernel also plays a role. Indeed, setting $W=0$ but using $GW$ QP energies, BSE yields real-valued excitation energies. It would be interesting to further investigate this issue by using the exact kernel together with $GW$ QP energies. This is left for future work.

\subsubsection{Correlation energy}
For the Hubbard dimer, we have $\EHF = -2t+U/2$, and the correlation energy given in Eq.~\eqref{eqn:intAC} can be calculated analytically. 
After a lengthy but simple derivation, one gets
\begin{widetext}
\begin{equation}
\resizebox{1\textwidth}{!}{$ \begin{split}
	\EcACBSE 
	& = - \frac{U}{2} 
	+ \frac{t^2-2U^2}{2U(2t+3U)} \qty{ \Delta\eps^{GW}-\frac{1}{2(t+U)}\sqrt{[-U^2+2(t+U)\Delta\eps^{GW}] \qty[U(2t+3U)+2(t+U)\Delta\eps^{GW}]} }\nonumber	\\
	& - \frac{t+2U}{2\sqrt{U(2t+3U)}} \qty( \frac{3t+4U}{2t+3U}+\frac{t}{U} ) \Delta\eps^{GW}
	\atan\qty{ -\frac{U\sqrt{U(2t+3U)}}{2\Delta\eps^{GW}(t+U)+\sqrt{[-U^2+2(t+U)\Delta\eps^{GW}][U(2t+3U)+2(t+U)\Delta\eps^{GW}]}}}.
	\end{split}$}
\label{eqn:EcanalAC}
  \end{equation}
  \end{widetext}

\begin{figure*}
	\includegraphics[width=0.33\textwidth]{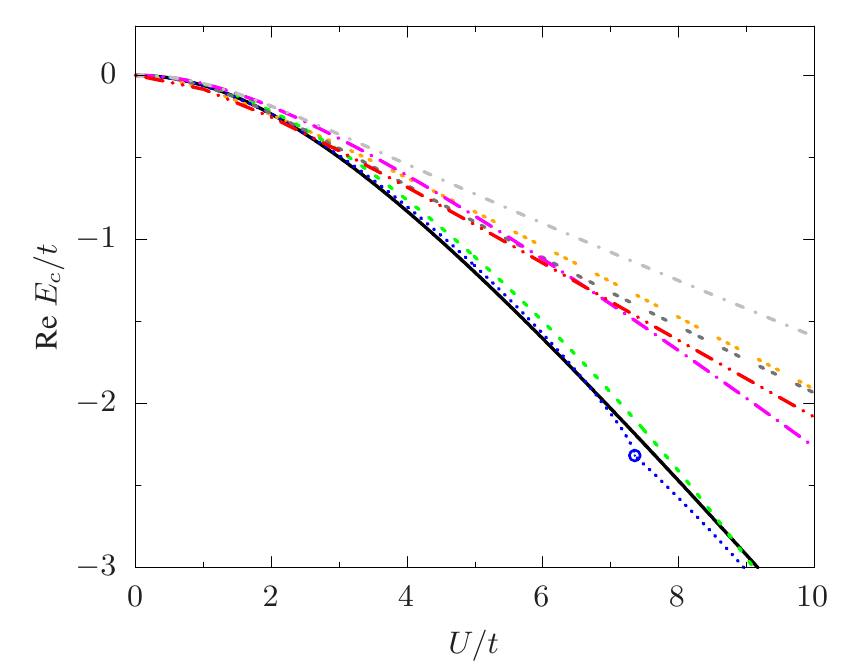}
    \includegraphics[width=0.33\textwidth]{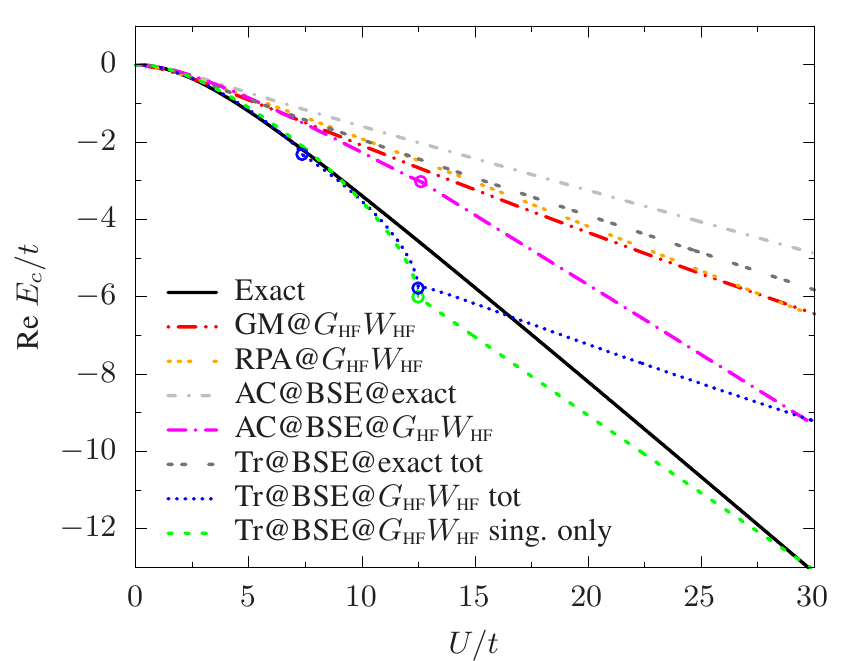}
    \includegraphics[width=0.33\textwidth]{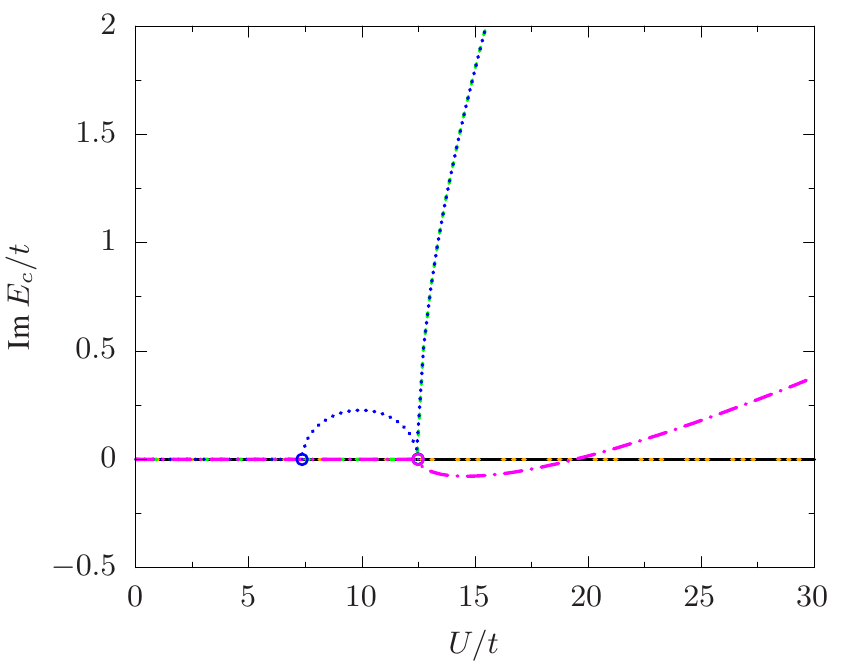}
 \caption{Real and imaginary parts of the BSE@$G_\text{HF}W_\text{HF}$ correlation energy as a function of $U/t$ at various levels of theory: total (dotted blue line) and singlet-only (dashed green line) Tr@BSE, AC@BSE (dot-dashed magenta line), RPA (triple-dotted orange line), GM (double-dot-dashed red line), and exact (solid black line). For comparison also the BSE@exact (Tr@BSE, double-dotted dark grey line ; AC@BSE, dot-dashed light grey line) correlation energies are shown. Discontinuities in the first derivative of the energy (corresponding to the appearance of complex poles) are indicated by open circles.}
\label{fig:cenergy}
\end{figure*}

Results are reported in Fig.~\ref{fig:cenergy} and are compared with the exact correlation energy \citep{Romaniello_2009a}
\begin{equation}
	\Ec = - \frac{\sqrt{16 t^2 + U^2}}{2} + 2t.
\end{equation}
The AC@BSE correlation energy does not possess the correct asymptotic behavior for small $U$, as Taylor expanding Eq.~\eqref{eqn:EcanalAC} for small $U$, we obtain 
\begin{equation}
	\EcACBSE = -\frac{U^2}{32t} - \frac{5U^3}{96t^2} + \frac{323 U^4}{6144t^3} + \order{U^4},
\end{equation}
while for the exact correlation energy behaves as
\begin{equation}
	\Ec = - \frac{U^2}{16t} + \frac{U^4}{1024t^3} + \order{U^6}.
\end{equation}
Moreover, we found that the radius of convergence of the small-$U/t$ expansion of $\EcACBSE$ is very small due to a square-root branch point for $U/t \approx -2/3$.

In the case of the trace formula \eqref{Eqn:tracef}, the singlet and triplet contributions behave as
\begin{subequations}
\begin{align}
	\EcTr{\singlet} & = - \frac{U^2}{32t} - \frac{7U^3}{128t^2} + \frac{99U^4}{2048t^3} + \order{U^5},
	\\
	\EcTr{\triplet} & = - \frac{U^2}{32t} + \frac{7U^3}{128t^2} - \frac{157U^4}{2048t^3} + \order{U^5},
\end{align} 
\end{subequations} 
which guarantees the correct asymptotic behavior for the total Tr@BSE correlation energy 
\begin{equation}
	\EcTrBSE = - \frac{U^2}{16t} - \frac{29U^4}{1024t^3} + \order{U^5},
\end{equation}
and cancels the cubic term (as it should).

The trace formula is strongly affected by the appearance of the imaginary excitation energies: as shown in Fig.~\ref{fig:cenergy} where we plot the real and complex components of the BSE@$G_\text{HF}W_\text{HF}$ correlation energy as functions of $U/t$ at various levels of theory, irregularities (\ie, discontinuities in the first derivative of the energy) appear at the values of $U/t$ for which the triplet and singlet energies become purely imaginary.  The ACFDT expression, instead, is more stable over the range of $U/t$ considered here with only a small cusp on the energy surface at the singlet instability point after which the real part of $\EcACBSE$ behaves linearly with respect to $U/t$.
Overall, however, the correlation energy obtained by the trace formula is almost on top of its exact counterpart over a wide range of $U/t$, with a rather small contribution from the triplet component, \ie, $\abs*{\EcTr{\triplet}} \ll \abs*{\EcTr{\singlet}}$. 
For comparison purposes, the RPA correlation energy, which is obtained from the trace or ACDFT formula using BSE@$G_{\text{HF}}W_{\text{HF}}$ with $W=0$ in the BSE kernel, is also reported in Fig.~\ref{fig:cenergy}. Both formulas yield the same correlation energies as expected, and they show no irregularities thanks to the fact that BSE excitation energies are real-valued at the RPA level. Also correlation energies obtained using BSE@exact (also shown in Fig.~\ref{fig:cenergy}) do not show irregularities for the same reason. Moreover, they show a visible upshift with respect to the corresponding  AC@BSE@$G_{\text{HF}}W_{\text{HF}}$ and Tr@BSE@$G_{\text{HF}}W_{\text{HF}}$ results, which worsens the agreement with the exact correlation energy. Finally, we observe that both expressions for the correlation energy (at BSE@$GW$ level) produce better results than the Galinski-Migdal formula \eqref{Eqn:GM}, as one can see from Fig.~\ref{fig:cenergy}, in particular at large $U/t$.

\section{Conclusions}
\label{sec:Conclusions}

In this work we have used the symmetric Hubbard dimer to better understand some features of the $GW$ approximation and of BSE@$GW$. In particular, we have found that the unphysical discontinuities that may occur in quasiparticle energies computed using one-shot or partially self-consistent $GW$ schemes disappear using full self-consistency. However, full self-consistency does not give an overall improvement in term of accuracy and, at least for the Hubbard dimer, $G_{\text{HF}}W_{\text{HF}}$ is to be preferred. 

We have also analyzed the performance of the BSE@$GW$ approach for neutral excitations and correlation energies. We have found that, at any level of self-consistency, the excitation energies become complex for some critical values of $U/t$. This seems related to the approximate nature of the $GW$ quasiparticle energies, since using exact quasiparticle energies (hence the exact fundamental gap) solves this issue. The BSE excitation energies are good approximations to the exact analogs only for a small range of $U/t$ (or $U/t\lesssim 2$ for the lowest singlet-singlet transition and $U/t\lesssim 6$ for the singlet-triplet transition), while the strong-correlation regime remains a challenge. 

The correlation energy obtained from these excitation energies using the trace (or plasmon) formula has been found to be in very good agreement with the exact results over the whole range of $U/t$ for which these energies are real. The occurrence of complex singlet and triplet excitation energies shows up as irregularities in the correlation energy. The ACFDT formula, instead, is less sensitive to this. However, we have found that the AC@BSE correlation energy is less accurate than the ones obtained using the trace formula. Both, however, perform better than the standard Galitski-Migdal formula. Finally, we have studied the small-$U$ expansion of the correlation energy obtained with the trace and ACFDT formulas and we found that the former, contrary to the latter, has the correct behavior when one includes both the singlet and triplet energy contributions. Our findings point out to a possible fundamental problem of the AC@BSE formalism. 

Although our study is restricted to the half-filled Hubbard dimer, some of our findings are transferable to realistic (molecular) systems. In particular: (i) a fully self-consistent solution of the $GW$ equation cures the problem of multiple QP solutions, avoiding in the process the appearance of discontinuities in key physical quantities such as total or excitation energies, ionization potentials, and electron affinities; (ii) a ``bad'' starting point ($G_0$ in the case of the Hubbard dimer) may result in the appearence of multiple QP solutions; (iii) potential energy surfaces computed with the trace formula and within the ACFDT formalism may exhibit irregularities due to the appearence of complex BSE excitation energies; (iv) for the Hubbard dimer at half-filling, the trace formula has the correct asymptotic behavior (thanks to the inclusion of singlet and triplet excitation energies) for weak interaction, contrary to its ACFDT counterpart. It would be interesting to check if it is also the case in realistic systems.

\section*{Conflict of Interest Statement}

The authors declare that the research was conducted in the absence of any commercial or financial relationships that could be construed as a potential conflict of interest.

\section*{Author Contributions}
All authors listed have made a substantial, direct, and intellectual contribution to the work and approved it for publication.

\section*{Funding}
This study has been partially supported through the EUR grant NanoX no ANR-17-EURE-0009 in the framework of the ``Programme des Investissements d'Avenir'' and by the CNRS through the 80$\,|\,$Prime program. PR and SDS also thank the ANR (project ANR-18-CE30-0025) for financial support.
PFL also thanks the European Research Council (ERC) under the European Union's Horizon 2020 research and innovation programme (grant agreement no.~863481) for financial support.


\appendix

\section{$G_0W_0$ equations for the half-filled Hubbard dimer}
\label{app:G0W0}
Starting from $G_0$, which reads
\begin{equation}
G_{0,IJ}(\omega)=\frac{1}{2}\qty[\frac{1}{\omega+t-\I\eta}+ \frac{(-1)^{(I-J)}}{\omega-t+\I\eta} ],
\label{Eqn:G_0-2e}
\end{equation}
one obtains the same $P$ and $W$ given in Eqs~\eqref{Eqn:P_site}-\eqref{Eqn:W_site}, from which the $G_0W_0$ self-energy reads
\begin{equation}
\label{Eqn:Sig_G0W0}
    \Sigma_{IJ}(\omega) =\frac{U}{2}\delta_{IJ}+\frac{U^2t}{2h}
    \left[ \frac{1}{\omega-(t+h)+\I\eta}  +\frac{(-1)^{I-J}}{\omega+(t+h)-\I\eta}\right].
\end{equation}
One then obtains the following $G_0W_0$ removal/addition energies 

\begin{subequations}
\begin{align}
	\eps_{1,\pm} & = + \frac{h}{2} +  \frac{U}{4} \pm \frac{\sqrt{
	(h+2t-U/2)^2 + 4 t U^2/h}}{2},
	\label{Eqn:QP_G0W0_b}
	\\
	\eps_{2,\pm} & =   -\frac{h}{2} +  \frac{U}{4} \pm \frac{\sqrt{
	(h+2t+U/2)^2+ 4 t U^2/h}}{2},
	\label{Eqn:QP_G0W0_a}
\end{align}
\end{subequations}
with the quasiparticle solutions being $\eps_\bn^\text{QP}=\eps_{1,-}$ and $\eps_\an^\text{QP}=\eps_{2,+}$, which correspond to the bonding and antibonding energies, respectively. The corresponding renormalization factors read
\begin{subequations}
\begin{align}
Z^{\text{QP}}_{\bn}=\frac{1}{2}+\frac{h+2t-\frac{U}{2}}{2\sqrt{
	(h+2t-U/2)^2 + 4 t U^2/h}},
	\label{Eqn:Z_QPb_G0W0}
	\\
	Z^{\text{QP}}_{\an}=\frac{1}{2}+\frac{h+2t+\frac{U}{2}}{2\sqrt{
	(h+2t+U/2)^2 + 4 t U^2/h}},
	\label{Eqn:Z_QPa_G0W0}
\end{align}
\end{subequations}
and $Z^{\text{sat}}_{\bn/\an}=1-Z^{\text{QP}}_{\bn/\an}$. We notice that the removal/addition energies and corresponding intensities given in Eqs (\ref{Eqn:QP_G0W0_b})-(\ref{Eqn:Z_QPa_G0W0}) correspond to the expressions (35) and (43)-(44) given in Ref.~\citep{Hellgren_2015} upon setting the nearest neighbour interaction $U_1$ to zero and $\epsilon_{H/L}=\pm t$.
\section{Numerical ev$GW$ and sc$GW$ calculations}
\label{app:meromorphic}

Following Ref.~\citep{vonFriesen_2010}, we see from Eq.~\ref{Eqn:spectralG} that the matrix elements of the exact Green's function $G$ in a generic orbital basis set can be expressed in the frequency domain as a sum over poles, \ie,
\begin{equation}
\label{Eqn:spectralGmero}
	G_{ij}(\omega) = \sum_\nu \frac{ G_{ij,\nu} }{ \omega - \eps_\nu + \I \eta \, \text{sgn}(\eps_\nu - \mu ) },
\end{equation}
where we introduced the spectral intensities
\begin{equation}
G_{ij,\nu}=\int d\bx_1 \bx_2 \phi^*_i(\bx_1) \psi_\nu(\bx_1) \psi^*_\nu(\bx_2) \phi_j(\bx_2).
\end{equation}
This representation remain valid for approximate Green's function, such as the non-interacting $G$ or its mean-field versions.
Likewise, $\Sigma$, $W$, and $P$ have similar representations.
Equation \eqref{Eqn:spectralGmero} allows us to evaluate convolutions and cross-correlations analytically.
Given the two functions 
\begin{subequations}
\begin{align}
\label{Eqn:spectralAmero}
	A(\omega) & = \sum_\nu \frac{ A_\nu }{ \omega - a_\nu + \I \eta \, \text{sgn}(a_\nu - \mu ) },
	\\
\label{Eqn:spectralBmero}
	B(\omega) & = \sum_\nu \frac{ B_\nu }{ \omega - b_\nu + \I \eta \, \text{sgn}(b_\nu - \mu ) },
\end{align}
\end{subequations}
their cross correlation functions
\begin{subequations}
\begin{align}
    C(\omega) & =\int \frac{d\omega'}{2\pi \I} A(\omega')B(\omega+\omega'),
    \label{eqn:cross+}
    \\
    D(\omega)& = \int \frac{d\omega'}{2\pi \I} A(\omega')B(\omega-\omega'),
    \label{eqn:cross-}
\end{align}
\end{subequations}
can be written as
\begin{align}
\begin{split}
    C(\omega) 
     =& -\sum_{b_\nu<\mu}\sum_{a_\xi>\mu} \frac{ A_\xi B_\nu}{ \omega - (b_\nu-a_\xi) - \I \eta }
    \\
    & + \sum_{a_\nu<\mu}\sum_{b_\xi>\mu} \frac{ A_\nu B_\xi}{ \omega - (b_\xi-a_\nu) + \I \eta },
\end{split}
    \\
\begin{split}
    D(\omega)
     =& \sum_{a_\nu<\mu}\sum_{b_\xi<\mu} \frac{ A_\nu B_\xi}{ \omega - (a_\nu+b_\xi) - \I \eta }
    \\
    & -\sum_{a_\nu>\mu}\sum_{b_\xi>\mu} \frac{ A_\nu B_\xi}{ \omega - (a_\nu+b_\xi) + \I \eta }.
\end{split}
\end{align}
Equations \eqref{eqn:cross+} and \eqref{eqn:cross-} enter, for example, in the evaluation of $P$ and $\Sigma$.

The Dyson equation for $G$ can then be solved in two steps: i) finding the poles of $G$, $\epsilon^{GW}_{i,\nu}$, which correspond to the zeroes of Eq.~\eqref{Eqn:QP} with $\phi_i(\br)=\phi_{\bn/\an}(\br)$, with, for example, an ordinary root finding algorithm; ii) once the positions of the poles are known, one can then compute the corresponding spectral weights via Eq.~\eqref{Eqn:Z}.

\bibliography{bibliography}

\end{document}